\documentclass[runningheads]{llncs}
\usepackage{graphicx}
\usepackage{hyperref}
\usepackage{todonotes}
\usepackage{amsmath}
\usepackage{amssymb}
\usepackage{physics}
\usepackage{adjustbox}
\usepackage[
    style=nature
]{biblatex}
\begin{document}

\title{Brain-Shift: Unsupervised Pseudo-Healthy Brain Synthesis for Novel
    Biomarker Extraction in Chronic Subdural Hematoma}
\titlerunning{Pseudo-Healthy Brain Synthesis in cSDH}

\author{Baris Imre \inst{1}, Elina Thibeau-Sutre \inst{1}, Jorieke Reimer
    \inst{3}, Kuan Kho \inst{2, 3}, Jelmer M. Wolterink \inst{1}}
\authorrunning{B. Imre et al.}

\institute{Department of Applied Mathematics, Technical Medical Center,
    University of Twente, Enschede, The Netherlands \and Clinical
    Neurophysiology
    Group, Techmed Centre, University of Twente, Enschede, The Netherlands
    \and
    Department of Neurosurgery, Medisch Spectrum Twente, Enschede, The
    Netherlands}

\maketitle

\begin{abstract}
    Chronic subdural hematoma (cSDH) is a common neurological condition
    characterized by the accumulation of blood between the brain and the dura
    mater. This accumulation of blood can exert pressure on the brain,
    potentially
    leading to fatal outcomes. Treatment options for cSDH are limited to
    invasive
    surgery or non-invasive management. Traditionally, the midline shift,
    hand-measured by experts from an ideal sagittal plane, and the hematoma
    volume
    have been the primary metrics for quantifying and analyzing cSDH. However,
    these approaches do not quantify the local 3D brain deformation caused by
    cSDH.
    We propose a novel method using anatomy-aware unsupervised diffeomorphic
    pseudo-healthy synthesis to generate brain deformation fields. The
    deformation
    fields derived from this process are utilized to extract biomarkers that
    quantify the shift in the brain due to cSDH.  We use CT scans of 121
    patients
    for training and validation of our method and find that our metrics allow
    the
    identification of patients who require surgery. Our results indicate that
    automatically obtained brain deformation fields might contain prognostic
    value
    for personalized cSDH treatment. Our implementation is available on:
    \url{github.com/Barisimre/brain-morphing}

\end{abstract}

\section{Introduction}

\begin{figure}[t!]
    \centering
    \includegraphics[width=\textwidth]{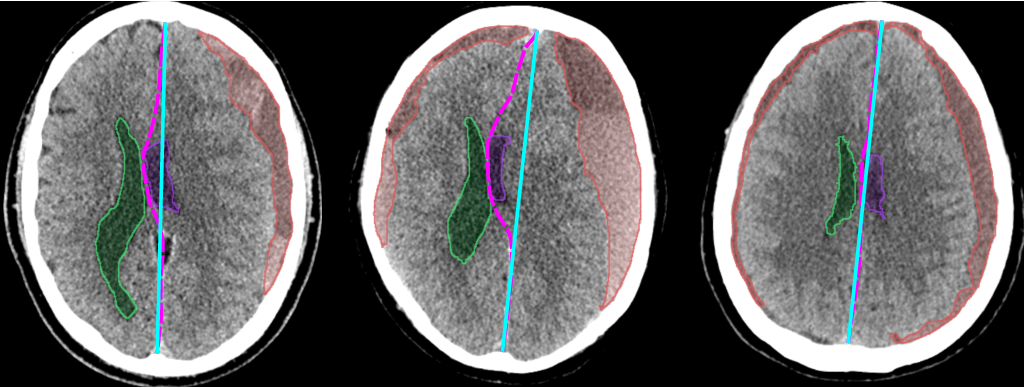}
    \caption{CT scans of cSDH patients selected for surgical intervention,
        showing the ideal midline (blue), shifted midline (pink), cSDH (red),
        left
        ventricle (purple), and right ventricle (green). Left: a unilateral
        hematoma
        with midline shift. Center: a bilateral hematoma with midline shift.
        Right: a
        bilateral hematoma without midline shift.}
    \label{fig:samples}
\end{figure}

With an aging population, chronic subdural hematoma (cSDH) is an increasingly
common neurological condition. cSDH is characterized by blood accumulation
between the brain and dura mater that can pressure the brain, deform it, and
potentially be fatal \cite{intro1}. CT visualizes hematomas well (Fig.
\ref{fig:samples}) and remains the most commonly used diagnostic tool for cSDH.
However, currently, there is no consensus on cSDH treatment following CT
imaging \cite{Mehta2018}. Treatment varies between wait-and-scan management and
surgery \cite{treatment}. Medical or endovascular treatments are subject to
large trials \cite{surgeryvswait}.

Clinical decision-making could be improved with objective metrics to quantify
the severity of cSDH. Currently, cSDH severity is quantified using geometric
properties such as the midline or midplane shift of the brain and the volume of
the hematoma ~\cite{heamtomats_volume_andshift, into_mls_prediction}. Various
methods have studied automatic determination of midline shift \cite{mls,
    Xiao2010, Nag2021, RuizheLiu2009} and midplane shift
\cite{xia2021automated},
and recent works have shown that the hematoma can be automatically segmented
using deep learning \cite{seg1}. However, these methods are global and fail to
take the local effects of cSDH into account. Moreover, in the case of symmetric
(bilateral) cSDH, midline shift might be minimal, while cSDH severity is
significant (Fig. \ref{fig:samples}).

In this paper, we introduce a novel approach to quantitatively assess the
severity of cSDH, by estimating the brain shift caused by the hematoma. We
estimate this deformation using an unsupervised deep learning-based
diffeomorphic method \cite{Ashburner_2007, dalca}. Unlike common image
registration problems, we do not have paired images for training our model
since pre-disease brain scans are unavailable for most patients. Instead, we
regularize our deformation model with personalized geometric priors to obtain a
pseudo-healthy, symmetric brain. All geometric priors, i.e., the ideal midplane
of the brain, 3D segmentation of hematomas and brain ventricles, and brain
masks, are automatically obtained.  We demonstrate that the diffeomorphic
deformation fields generated by our methodology, and biomarkers extracted from
these deformation fields, offer an accurate assessment of cSDH severity that
complements traditional midline shift metrics.

\section{Materials and Methods}
\begin{figure}[t]
    \centering
    \includegraphics[width=\textwidth]{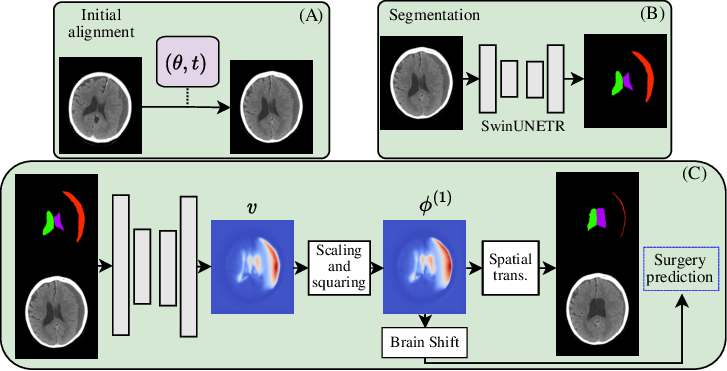}
    \caption{Overview of our process. Images are aligned to a sagittal plane
        (A) and automatically segmented (B). The CT scan and segmentations are
        used as
        input to a deep diffeomorphic registration model (C). Deformation
        fields
        provided by this model are used to distinguish between patients
        requiring
        surgery by various biomarkers. }
    \label{fig:overview}
\end{figure}

Figure~\ref{fig:overview} outlines our method. Given a CT scan of a brain with
cSDH, we automatically align the brain with canonical axes. Then, a Swin UNETR
\cite{swinunetr} model automatically segments the hematomas and ventricles.
Images and segmentation masks are jointly used in an unsupervised diffeomorphic
VoxelMorph-based \cite{dalca} model that provides local velocity fields
required to virtually remove the hematoma and translate the brain into its
pseudo-healthy version.

\subsection{Data}
We used a retrospectively collected dataset of 121 patients scanned at
Medisch Spectrum Twente between 2011 and 2019. A waiver from the local
medical ethics committee was obtained. We excluded all scans that showed cSDH
after a first surgical intervention. Selected scans had an in-plane resolution
of $0.41 \pm 0.065 \times 0.40 \pm 0.070$ mm and a slice thickness of $3.42 \pm
    0.64$ mm. All scans were resampled to a spatial resolution of $0.40 \times
    0.40
    \times 1.50$ mm. For the majority of patients, we know whether - based on
the
CT scan and other parameters - surgery was performed to mitigate the effects of
cSDH. In total, 46 patients received surgery, and 68 did not. Seven patients
without surgical status were excluded from the dataset. Surgical status serves
as the endpoint for our prediction model. In a subset of 25 scans, the
hematoma, left ventricle, and right ventricle were manually annotated in 3D
using the XNAT OHIF viewer \cite{xnat}. Annotations were made by technical
physicians in training, and verified by a neurologist to ensure the quality and
accuracy of the annotations. Moreover, manual measurements were obtained of the
midline shift of each subject within the dataset.

\subsection{Brain Symmetry Losses}
\label{losses}
In the initial alignment and diffeomorphic registration steps of our algorithm
((A) and (C) in Fig.~\ref{fig:overview}), we use two fully differentiable
losses that quantify brain symmetry. We initially split each volume $X \in
    \mathbb{R}^3$ into two equally sized volumes ($X_{left}$ and $X_{right}$)
by
the mid-sagittal plane, halving the scan perfectly into two. Both of the losses
use these pseudo-hemispheres as input. The two losses are the SSIM loss,
derived as the negative of the structural symmetry index (SSIM), \cite{ssim}
$\mathcal{L}_{SSIM}$ and Jeffreys divergence loss $\mathcal{L}_{jeffrey}$
\cite{Jeffreys1998-zj}. Let $H: \mathbb{R}^3 \rightarrow \mathbb{R}^1$ be the
differentiable histogram function with $n$ bins, introduced by Ustinova et al.
\cite{diff_hist}. The Jeffreys divergence loss $\mathcal{L}_{jeffrey}$ is
calculated as the following, using histograms as probability distribution
function surrogates:

\begin{align}
    \mathcal{L}_{jeffrey} = \sum_n H_n(X_{left}) \log
    \frac{H_n(X_{left})}{H_n(X_{right})} + \sum_n H_n(X_{right}) \log
    \frac{H_n(X_{right})}{H_n(X_{left})}
    \label{eq_jeffrey}
\end{align}

A higher value for $\mathcal{L}_{jeffrey}$ indicates that the two
pseudo-hemispheres being compared have different intensity distributions,
either due to a misalignment from the mid sagittal plane or the presence of a
hematoma, contributing to a collection of many similar voxels grouped in
intensity.

\subsection{Initial Alignment and Segmentation}
We rigidly align 3D volumes to establish a symmetry axis on the sagittal plane
(the ideal midplane or mid-sagittal plane). In other words, we want the brains
to be as symmetric as possible with respect to the center (mid) sagittal plane.
We use iterative optimization with Adam \cite{adam}. At each step, we calculate
the combined symmetry loss explained in Section \ref{losses} together with a
volume loss term. We optimize four variables per scan, corresponding to the
pitch, yaw, roll ($\theta$), and translation ($t$) of the 3D volume, similarly
to the process described by Prima et. al \cite{centering_2002}. The volume loss
ensures that both of the halves in the results have equal volumes of
foreground, counting the total volume in both halves, using a differentiable
binary operator $B: \mathbb{R}^3 \rightarrow \mathbb{R}^3$ that maps any non
zero items to one and every zero items to zero. We find this volume loss
especially useful in cases where the head is located far from the center of the
scan.

\begin{align}
    \mathcal{L}_{volume} = \frac{\left| \sum B(X_{left}) - \sum B(X_{right})
        \right|}{\sum B(X)}
\end{align}

We segment the subdural hematomas, left ventricles, and ventricles in aligned
3D CT scans using a Swin UNETR \cite{swinunetr} trained on manual
segmentations. During training, we augment our data using random rotations,
random 3D crops, and sagittal flips. We denote the result of segmenting image
$X$ as $S(X)$ below, with each segmentation class denoted as $S_{class}$.

\subsection{Diffeomorphic Pseudo-Healthy Synthesis}
For any brain with cSDH, we predict the deformation required to translate it
into a pseudo-healthy equivalent. We train an image-to-image model that, given
input volumes and masks, estimates a velocity field $\mathit{v}$ that can be
integrated over $t = [0, 1]$ using scaling and
squaring~\cite{scalingandsquaring} to obtain a deformation field $\phi^{(1)}$.
This field is applied to the CT scan to obtain the pseudo-healthy brain
$\phi(X)$.

We minimize a compound loss term, based on the pseudo-healthy brain $\phi(X)$,
the deformation field ($\phi^{(1)}$), and the deformed segmentation masks
($\phi(S(X))$). We aim to maximize the internal symmetry of the pseudo-healthy
brains with the loss terms explained in Section \ref{losses}. We separately
apply the deformation to the segmentation masks ($S_{left}(X)$, $S_{right}(X)$,
and $S_{hematoma}(X)$) and calculate two loss terms. We maximize the symmetry
of the ventricles by optimizing for their overlap when one of them is reflected
on the sagittal plane using a Dice loss (Eq. \ref{ventricle}). We target the
removal of the hematomas by optimizing to reduce the volume in the deformed
hematoma mask formalized with Eq. \ref{hematoma}.

\begin{align}
    \mathcal{L}_{ventricle} & = Dice(\phi(S_{left}(X)), \;
    SagittalFlip(\phi(S_{right}(X)))) \label{ventricle}               \\
    \mathcal{L}_{hematoma}  & = 1 - \frac{\sum S_{hematoma}(X) - \sum
        \phi(S_{hematoma}(X))}{\sum S_{hematoma}(X)} \label{hematoma}
\end{align}

We ensure the preservation of the skull with $\mathcal{L}_{skull}$, using the
Dice loss of the original and deformed skull. We regularize the registration
methods by introducing two additional loss terms. Firstly, we limit the
shrinkage of non-hematoma voxels by a Jacobian regularizer. The Jacobian
determinant $\text{det} \nabla \Phi$ at any location indicates the expansion or
shrinkage. We aim to regularize the Jacobian of the non-hematoma voxel to 1 and
the hematoma voxels to 0. Secondly, we approximate the spatial gradient
differences of the velocity field with the same method as Balakrishnan et. al.
\cite{Balakrishnan_2019}, shifting in all three dimensions by one and
subtracting the velocity magnitudes. We find that applying this loss to the
velocity field $\mathit{v}$ rather than the deformation field $\phi^{(1)}$
produces smoother results. The deformation loss $\mathcal{L}_{deformation}$ is
defined in Eq. \ref{eq:deformation}.

\begin{equation}
    \label{eq:deformation}
    \begin{gathered}
        \mathcal{L}_{deformation} = \;	\lambda_1 \; \mathcal{L}_{jeffrey} +
        \lambda_2 \; \mathcal{L}_{ssim} + \lambda_3 \; \mathcal{L}_{ventricle}
        +
        \lambda_4 \; \mathcal{L}_{hematoma} \\
        + \lambda_5 \; \mathcal{L}_{skull} +\lambda_6 \; \mathcal{L}_{jacobian}
        +
        \lambda_7 \; \mathcal{L}_{gradient}  \\
    \end{gathered}
\end{equation}

\subsection{cSDH Severity Prediction}
\label{brain-shift}
We use the deformation fields generated per scan to calculate three novel
biomarkers of brain deformation, namely, the maximum, average, and sum of the
magnitudes of voxel-wise deformation vectors (in mm). We divide the markers
into two categories, conventional markers including hematoma volume and midline
shift, and our three novel biomarkers.
We fit a logistic regression classifier for each biomarker, in each data
subset, with surgical outcome as target. We evaluate the model's performance in
the identification of patients requiring surgery using ROC curves. Moreover, we
utilize TPOT \cite{tpot} to identify the best-performing model for prediction
based on combinations of these derived markers.

\section{Experiments and Results}

All experiments were conducted on a single machine with 128 GBs of RAM and an
NVIDIA L40 GPU (48 GBs VRAM). Methods were implemented in PyTorch \cite{torch},
and code is provided on GitHub\footnote{github.com/Barisimre/brain-morphing}. We
perform a five-fold cross-validation of the full model.

\begin{figure}[t]
    \centering
    \includegraphics[width=\linewidth]{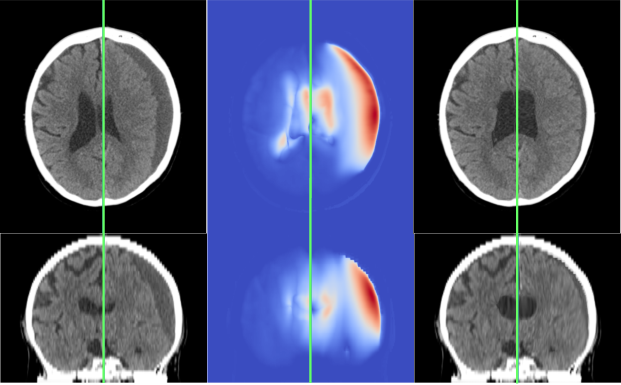}
    \caption{Axial and coronal slide visualizing registration results and
        mid-sagittal plane (green). From left to right: the original brain with
        a
        unilateral (visible on the right side) cSDH, the magnitude of the
        deformation
        field, and the resulting pseudo-healthy brain. The hematoma is mostly
        removed
        and ventricular symmetry is restored.}
    \label{fig:example}
\end{figure}

\subsection{Initial Alignment and Segmentation}
We observed that 150 iterations with a learning rate of 0.03 and Adam optimizer
\cite{adam} was sufficient for initial image alignment. The Swin UNETR model is
trained for 5000 epochs using the Adam optimizer and a learning rate of 3e-4.
We observed an average Dice score of 0.82 for hematomas and 0.86 for both of
the ventricles.

\subsection{Pseudo-Healthy Synthesis}
For the pseudo-healthy synthesis process, we used the Adam optimizer with a
learning rate of 3e-4. We trained this model for 20,000 iterations. Loss
coefficients in Eq. \ref{eq:deformation} were chosen as 5.0 for
$\mathcal{L}_{jacobian}$, $\mathcal{L}_{gradient}$, and $\mathcal{L}_{skull}$
and 1.0 for the rest.
Figure \ref{fig:example} shows a qualitative example where the brain was
clearly deformed in the hematoma area as well as on the contralateral side.
Across all images, we observed an average hematoma volume reduction of 73
percent with a standard deviation of 13 percent. This indicates that the method
is not able to remove all hematoma. In general, we found thin hematomas
circling the brain around more difficult to reduce due to the regularization of
the deformation fields.

\begin{figure}[t]
    \centering
    \includegraphics[width=\textwidth]{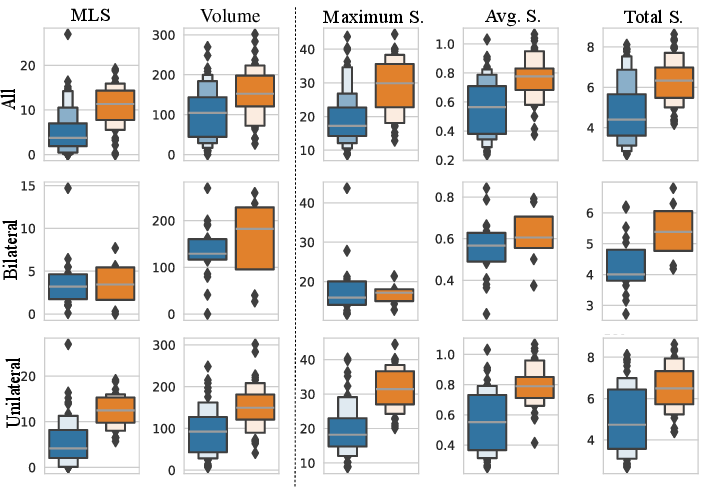}
    \caption{Box plots for five biomarkers for three subsets of our data,
        comparing the distributions of patients that required surgery (orange)
        compared
        to the rest (blue). The midline shift (MLS) is hand measured in mm.
        Hematoma
        volume (Volume) is inferred from the segmentation results in mm.
        Maximum,
        average, and total shift are calculated as explained in subsection
        \ref{brain-shift} in millimeters. Many of the biomarkers fail at
        distinguishing
        surgical status of bilateral hematomas.}
    \label{fig:boxen-plots}
\end{figure}

\subsection{Surgery Prediction}
We investigated the distribution of five biomarkers in the group of patients
that did not require surgery and the group that did. These biomarkers are the
midline shift, hematoma volume, and our three markers derived from the
deformation fields: maximum, average, and sum. We performed a subgroup analysis
on patients with bilateral or unilateral hematomas. Figure
\ref{fig:boxen-plots} confirms our hypothesis that midline shift (MLS) poorly
separates patients with bilateral cSDH. However, separating these cases is also
difficult for our maximum shift feature.

\begin{figure}[t]
    \centering
    \includegraphics[width=\textwidth]{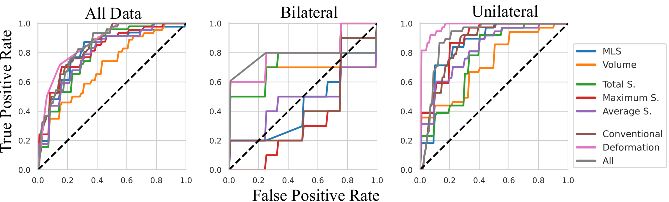}
    \caption{ROC curves of surgical status classifiers in either the full
        population (left) or subgroups of patients with bilateral (center) or
        unilateral (right) hematomas.}
    \label{fig:rocs}
\end{figure}

The receiver operating characteristic curves (ROC) for each of the
classification models is shown in Figure \ref{fig:rocs}. For the bilateral
subset, only the total shift and hematoma volume as single markers were able to
perform better than random in classification. Table \ref{tab:auc} shows the
area under the curve (AUC) results corresponding to these ROCs.
In each subset, we observed that a marker generated from the deformations
fields performs better than traditional markers, while the combination of all
deformation-based markers performed the overall best.

\begin{table}[t]
    \caption{AUC scores of all classification models. MLS = midline shift. We
        present models for individual conventional and deformation-based
        biomarkers and
        combinations of biomarkers.}
    \centering

    % \resizebox{\linewidth}{!}{
    \begin{tabular}{c|cc|ccc|ccc|}
                                                &
        \multicolumn{2}{|c|}{Conventional}      &
        \multicolumn{3}{|c|}{Deformation-based} & \multicolumn{3}{|c|}{Joint}
        \\
                                                & MLS
                                                & Volume                      & Average & Max    & Sum    & Convent. & Deform.
                                                & All
        \\ \hline
        All Data                                & $0.80$
                                                & $0.70$                      & $0.79$  & $0.81$ & $0.79$ & $0.84$   &
        $\mathbf{0.87}$                         & $0.84$
        \\
        Bilateral                               & $0.44$
                                                & $0.67$                      & $0.48$  & $0.37$ & $0.76$ & $0.42$   &
        $\mathbf{0.79}$                         & $0.77$
        \\
        Unilateral                              & $0.85$
                                                & $0.72$                      & $0.82$  & $0.87$ & $0.76$ & $0.86$   &
        $\mathbf{0.96}$                         & $0.90$
        \\ \hline
    \end{tabular}%
    % }

    \label{tab:auc}
\end{table}

\section{Discussion and Conclusion}
We have presented a novel pseudo-healthy brain synthesis process, with
applications in cSDH patients. We have shown that symmetry losses can be used
to obtain visually plausible pseudo-healthy CT scans of patients, and
deformation fields from which features can be extracted for personalized
treatment prediction. Our results confirm the hypothesis that biomarkers are
affected by the laterality of the cSDH, where most biomarkers underperform in
bilateral cSDH.

This study shows that there is valuable information in 3D context for cSDH
assessments. Moreover, our work highlights the value of machine learning for
cSDH and neurosurgery. First, we fully automatically segment cSDH and
ventricles in 3D CT volumes. Second, we obtain deformation fields with a
generalizable model. Third, we use machine learning models to identify patients
in need of surgery. The deformation fields that we obtain are diffeomorphic,
which means that they also represent the inverse process of cSDH formation. In
future works, this could be used to synthesize data of cSDH patients from
healthy brains, and further our understanding of cSDH formation.

Our study has limitations. First, ground truth labels for our outcome - surgery
or not - are subjective and vary per clinician \cite{variation_in_hostpitals},
especially given the current lack of objective criteria for this decision. In
future work, we aim to use our novel biomarkers to predict objective outcomes
and correlate with invasive measurements of pressure \cite{pressure_measure}.
Second, our model is not end-to-end, and errors might accumulate along our
pipeline. For example, imperfect alignment might affect the performance of the
segmentation model, which in turn might lead to incorrect deformation fields.
In future work, we will investigate the development of an end-to-end model for
joint segmentation and registration. Moreover, a larger data set might allow
the extraction of more informative biomarkers. It has been shown that
deformation fields might contain valuable prognostic
information~\cite{trebeschi2021development}. Finally, our heuristic symmetry
losses could be extended with data-driven perceptual losses that define what a
healthy brain looks like~\cite{baumgartner2018visual}.

In conclusion, our results indicate that automatically obtained brain
deformation fields might contain prognostic value for personalized cSDH
treatment.

\subsubsection{Acknowledgments} This work has been financially supported by Pioneers in Healthcare. Jelmer M. Wolterink was supported by the NWO domain Applied and Engineering Sciences Veni grant (18192).

\printbibliography

\end{document}